\documentclass[onecolumn]{autart} 

\usepackage{color,graphicx}     

\usepackage{lscape}
\usepackage{setspace}
\usepackage{amsmath,amsfonts,amssymb}
\usepackage{hyperref}
\usepackage[all]{hypcap}
\usepackage[square,numbers,sort&compress]{natbib}
\newtheorem{theorem}{theorem}[section]
\newtheorem{corollary}[theorem]{Corollary}

\begin{document}

\begin{frontmatter}

\title{Stochastic Nonlinear Dynamics of Interpersonal and Romantic Relationships}     
\thanks[footnoteinfo]{Corresponding author: A.~Cherif. Tel. +1-480-965-2115. Fax +1-480-727-7346.}

\author[MCMSC,SHES]{Alhaji Cherif\thanksref{footnoteinfo}}\ead{alhaji.cherif@asu.edu},    
\author[MCMSC,SHES]{Kamal Barley}\ead{kamal.barley@asu.edu}          

\address[MCMSC]{Mathematical and Computational Modeling Sciences Center, Arizona State University, Tempe AZ 85287\\ } 
\address[SHES]{School of Human Evolution and Social Change, Arizona State University, Tempe AZ 85287\\ } 

\begin{abstract}                          
Current theories from biosocial (e.g.: the role of neurotransmitters in behavioral features), ecological (e.g.: cultural, political, and institutional conditions), and interpersonal (e.g.: attachment) perspectives have grounded interpersonal and romantic relationships in normative social experiences. However, these theories have not been developed to the point of providing a solid theoretical understanding of the dynamics present in interpersonal and romantic relationships, and integrative theories are still lacking. In this paper, mathematical models are use to investigate the dynamics of interpersonal and romantic relationships, which are examined via ordinary and stochastic differential equations, in order to provide insight into the behaviors of love. The analysis starts with a deterministic model and progresses to nonlinear stochastic models capturing the stochastic rates and factors (e.g.: ecological factors, such as historical, cultural and community conditions) that affect proximal experiences and shape the patterns of relationship. Numerical examples are given to illustrate various dynamics of interpersonal and romantic behaviors (with emphasis placed on sustained oscillations, and transitions between locally stable equilibria) that are observable in stochastic models (closely related to real interpersonal dynamics), but absent in deterministic models.
\end{abstract}


\end{frontmatter}

\section{Introduction}
Interpersonal relationships appear in many contexts, such as in family, kinship, acquaintance, work, and clubs, to name a few. The manifestation of interpersonal relationships in society comes in many forms ranging from romantic, parent-child, friendships, comradeship, casual, friend-with-benefits, soul-mates, dating to more recently Internet relationships. The most intriguing of all of these interpersonal relationships, which is also a dominant phenomenon and fundamental in human social life and interaction, is romantic relationship (\cite{Furman2006}-\cite{Collins2003}).

Romantic relationships refer to the mutually ongoing interactions between two or more individuals.  Recent works show that romantic relationships are more common among adolescents than has previously been assumed, with more than half of adolescents in the United States being involved in some form of romantic relationships (\cite{Allen1999}-\cite{Laursen1997}).  More than 70\% of high school and college students report having had a special romantic relationship in the previous years, and also report more frequent interactions with romantic partners than with parents, siblings, and/or friends (\cite{Laursen1997}-\cite{Collins2006} ).  In the case of adult, the study of romantic behaviors may provide invaluable insight as to why majority of romantic relationships fail or do not make it to engagement and/or marriage (\cite{Lloyd1984}).  Surra and Hughes (\cite{Surra1997}) found that more than half (54\%) of couples in their studies exhibit unpredictable and nonlinear relational trajectories involving large number of turning and tipping points.  Partners identified events such as new rivals, unresolved differences, meeting partnerÕs family, and job changes as turning and tipping points that greatly changed and influenced the nature, quality and progress of the relationship.  Similar findings by others scholars have appeared in various literatures (\cite{Allen1999}-\cite{Laursen1997} and \cite{Brown2002}-\cite{Carver2003}).

Research on romantic relationships among adolescents has gained traction with emphases on the quality of relationships and their potential implications for positive and negative developmental and socio-psychological outcomes.  Theories in biosocial (e.g.: effect of neurotransmitters in behavioral features) (\cite{Bus2005}-\cite{Ellis2004}), ecological (e.g.: cultural, political, and institutional conditions) (\cite{Larson2004}), and interpersonal (e.g.: attachment) (\cite{Hinde1997}-\cite{Collins1999}) studies have grounded romantic relationships in normative social experiences and paradigms.  However, these theories have not been developed to the point of providing a solid theoretical understanding of the various dynamics present in romantic relationships; in addition, integrative theories are still lacking.  The study of relationships has begun to hold both the artistic imaginations and interdisciplinary intellectual interests of various scholars in the fields of sociology, biology, neuroscience, psychology, anthropology, and mathematics (\cite{Strogatz1988}-\cite{Liao2007}).  Since experiments in these areas are difficult to design and may be constrained by ethical considerations, mathematical models can play a vital role in studying the dynamics of relationships and their behavioral features.  However, there are few mathematical models capturing the various dynamics of romantic relationships.  In this paper, we study both deterministic and stochastic models relationship from interpersonal perspective.

Deterministic differential equations have been used extensively to study dynamic phenomena in a wide range of fields, ranging from physical, natural, biological to social sciences.  The mathematical models capturing the dynamics of love between two people have recently gained attention among many researchers (\cite{Rinaldi1998a}-\cite{Wauer2007}) who have provided extensions to Strogatz's seminal model.  In an one-page influential work (\cite{Strogatz1988}) and later in a book (\cite{Strogatz1994}), Strogatz applied a system of linear differential equations to study Shakespearean model of love affair of Romeo and Juliet. Rinaldi (\cite{Rinaldi1998a} and \cite{Rinaldi1998b}), Sprott (\cite{Sprott2004}),  Liao and Ran (\cite{Liao2007}) and Wauer et al (\cite{Wauer2007}) have investigated realistic perturbations and extensions of \textit{Strogatzian} model by including features such as attraction factor (\cite{Rinaldi1998a}, \cite{Rinaldi2004}, \cite{Liao2007}, and \cite{Wauer2007}), delay and nonlinear return functions (\cite{Wauer2007}), and three-body love affairs or love triangles (\cite{Sprott2004} and \cite{Wauer2007}).  Rinaldi investigated the three mechanisms of love dynamics: instinct, return, and oblivion in (\cite{Rinaldi2004}), making the model more realistic due to the fact that it accounts for the growth of feeling from a state of indifference.  In (\cite{Rinaldi1998a}), Rinaldi proposed a three dimensional model to describe the cyclical love dynamics of Laura and Patriarch and introduced nonlinear return and oblivion functions, and poems written by Patriarch are used to validate the dynamics. Gottman et al (\cite{Gottman2002}) employed discrete dynamical models to describe the interaction between married couples; Liao and Ran (\cite{Liao2007}) studied time delays, nonlinear coupling and Hopf bifurcation conditions. Recently, Wauer et al (\cite{Wauer2007}) examined various models, starting with a time-invariant two dimensional linear and nonlinear models and concluding with time-dependent fluctuations in the source-terms and parameters.  In previous papers (\cite{Rinaldi1998a}-\cite{Rinaldi1998b}, and \cite{Sprott2004}-\cite{Wauer2007}), only dyadic interactions are considered, and other effects such as personalities and differential appeals of the individuals are ignored. As a result, learning and adaptation processes are ruled out. In this paper, we investigate stochastic dynamical models.  But first, we summarized previous deterministic models (\cite{Strogatz1988}, \cite{Cherif2009} and \cite{Rinaldi1998a}) before developing an equivalent stochastic model.

\section{Models and Stability Analysis}
In this section, we study two models with two state variables.  The variables $X_{1}$ and $X_{2}$ are the measures of love of individual 1 and 2 for their respective partners, where positive and negative measures represent positive (e.g.: friendship, passionate, intimate) and negative (e.g.: antagonism and disdain) feelings, respectively.  We first propose a deterministic system of differential equations to model the dynamic of romantic relationship, and later extend the model naturally to a stochastic dynamic model, where the deterministic rates become the stochastic rates.  Using the typology of Strogatz (\cite{Strogatz1988}-\cite{Strogatz1994}) and Sprott (\cite{Sprott2004}), the four romantic styles are summarized in figure 1.

\begin{figure}
\begin{center}
\includegraphics[height=8.5cm]{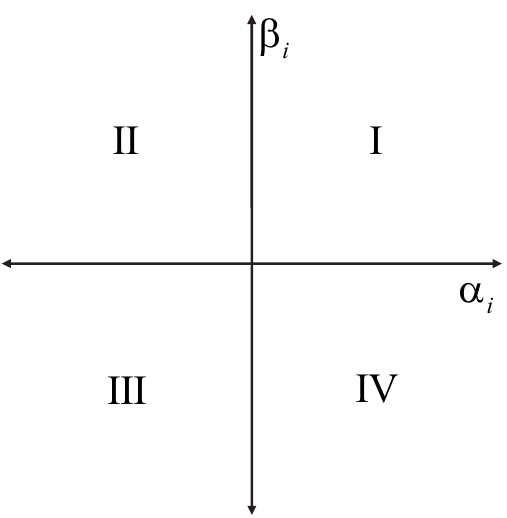}    
\caption{Typology and Characterization of Romantic Style}  
\label{fig1}                                 
\end{center}                                 
\end{figure}

\textbf{Summary of Figure 1:}

 \textit{\textbf{Region I}:} Eager Beaver: individual 1 is encouraged by his own feelings as well as that of individual 2 ($\alpha _{i} >0$ and $\beta _{i} >0$).\\
 \textit{\textbf{Region II}:} Secure or Cautious lover: individual 1 retreats from his own feelings but is encouraged by that of individual 2 ($\alpha _{i} <0$ and $\beta _{i} >0$).\\
\textit{\textbf{Region III}:} Hermit: individual 1 retreats from his own feelings and that of individual 2 ($\alpha _{i} <0$ and $\beta _{i} <0$).\\
\textit{\textbf{Region IV}:} Narcissistic Nerd: individual 1 wants more of what he feels but retreat from the feelings of individual 2 ($\alpha _{i} >0$ and $\beta _{i} <0$).\\

This classification allows us to characterize the dynamics exhibited by various combinations of a variety of romantic styles (see Fig. 1).  Previous papers (\cite{Strogatz1994}-\cite{Cherif2009}, \cite{Rinaldi1998a} and \cite{Sprott2004}) have considered various dynamics using all possible combinations in the sign of parameters $\alpha _{i} $ and $\beta _{i} $.  In this paper, we focus primarily on dynamics observed when individuals in region \textit{II} interact with individuals in region \textit{III}.  We later focus on the features of stochastic model that are not present in the deterministic dynamics (\cite{Cherif2009}).  For instance, Cherif (\cite{Cherif2009}) numerically showed that romantic relationships can exhibit exotic dynamics such as sustained oscillations while deterministic models show damped oscillations (e.g. stochastic resonance) and diffusion of trajectories between equilibria. In this section, we proceed as follows: deterministic (linear and nonlinear) models are provided, and are then followed by their stochastic equivalents. In the Deterministic Model section, we use standard deterministic linear romantic dynamic model (\cite{Strogatz1994}) and nonlinear model provided in (\cite{Cherif2009}).  In Stochastic Dynamic Models, stochastic versions of the models studied in section 2.1 are investigated with special emphasis on behaviors not observed in the deterministic dynamics. We then provide an extension, where differential parameters are considered.

\subsection{Deterministic Models}
 In (\cite{Cherif2009}), a more general model was proposed.  In this paper, we focus on dyadic relational dynamics characterized by \textit{regions} \textit{II} and \textit{III} interactions of figure 1 (secure or cautious love dynamics).  To model the behavioral features of romantic dynamics, the following deterministic model is proposed:
\begin{eqnarray} \label{subeqn:list}
\frac{dX_{1} }{dt} =-\alpha _{1} X_{1} +\beta _{1} X_{2} \left(1-\varepsilon X_{2} ^{2} \right)+A_{1} \label{subeqn:first} \\
\frac{dX_{1} }{dt} =-\alpha _{1} X_{1} \label{e1}+\beta _{1} X_{2} \left(1-\varepsilon X_{2} ^{2} \right)+A_{1} \label{subeqn:second}
\end{eqnarray}
 for $\left(X_{1} ,X_{2} \right)\in R\times R$, where $\alpha _{i} >0$ is non-negative, and $\beta _{i} $ and $A_{i} $ $i=1,2$ are real constant, respectively.  These parameters are oblivion, reaction and attraction constants, respectively.  For  $\beta _{i} $ and $A_{i} $, we relax positivity condition.  In the equations above (Eqs.~\ref{subeqn:first}-~\ref{subeqn:second}), we assume that feelings decay exponentially fast in the absence of partners.  The parameters specify the romantic style of individuals 1 and 2.  For instance, $\alpha _{i} $ describes the extent to which individual \textit{i} is encouraged by his/her own feeling.  In other words, $\alpha _{i} $  indicates the degree to which an individual has internalized a sense of his/her self-worth.  In addition, it can be used as the level of anxiety and dependency on other's approval in romantic relationships.  The parameters $\beta _{i} $ represent the extent to which individual $i$ is encouraged by his/her partner, and/or expects his/her partner to be supportive.  It measures the tendency to seek or avoid closeness in a romantic relationship.  Therefore, the term $-\alpha _{i} X_{i} $ say that the love measure of $i$, in the absence of the partner decay exponentially; and ${\raise0.5ex\hbox{$\scriptstyle 1 $}\kern-0.1em/\kern-0.15em\lower0.25ex\hbox{$\scriptstyle \alpha _{i}  $}} $ is the time required for love to decay.  We propose this structure as opposed to model described in previous literature (\cite{Rinaldi1998a}-\cite{Wauer2007}), because romantic relationships (as in any interpersonal relationship) are not linear, especially return factors. The functional structure of return factors are motivated by dynamics often portrayed in romance novels and the tragic outcomes illustrated in them (Shakespeare's \textit{Romeo and Juliet}, Patriarch's \textit{Canzoniere and Posteritati}, Tolstoy's \textit{Anna Karenina}, Flaubert's \textit{Madame Bovary}, more recently the tragic \textit{myspace} suicide of Megan Meier). The constant $\varepsilon $ in the return function can be interpreted as the compensatory constant. For example, the romantic dynamics between Laura Winslow and Steve Urkel in the sitcom Family Matter can serve as a popular media example.  When Steve Urkel despairs, Laura Winslow feels sorry for him and her antagonism is overcome by feeling of pity. As a result, she reverses her reaction to passion. This behavioral characteristic is captured by the function of reaction or return function (e.g.: $\beta_{1} X_{2} \left(1-\varepsilon X_{2}^{2} \right)$). This expression captures the compensation for antagonism with flattery, or pity, for positive and negative values of $X_{2} $ in $\beta_{1}  X_{2}\left(1-\varepsilon X_{2}^{2} \right)$, respectively. For $\varepsilon =0$, the model reduces to the models proposed by Strogatz (\cite{Strogatz1988}-\cite{Strogatz1994}), and others (\cite{Rinaldi1998a}-\cite{Liao2007} and \cite{Sprott2005}-\cite{Wauer2007}).  In \textit{Strogatzian} model of love affair, which corresponds to the case where  $\varepsilon = 0$ in equations 1 and 2, the equilibrium point, $\left(\bar{X}_{1} ,\bar{X}_{2} \right)$,  is satisfied by the following equations (\cite{Guckenheimer1983}):

\begin{eqnarray}\label{e2}
\bar{X}_{1} =\frac{\alpha _{2} A_{1} +\beta _{1} A_{2} }{\alpha _{1} \alpha _{2} -\beta _{1} \beta _{2}} \\
\bar{X}_{2} =\frac{\alpha _{1} A_{2} +\beta _{2} A_{1} }{\alpha _{1} \alpha _{2} -\beta _{1} \beta _{2}}
\end{eqnarray}

For such a system ($Strogatzian$ Model),  we note the following theorem:
\begin{thm}
The equilibrium $\left(\bar{X}_{1}, \bar{X}_{2} \right)$ is non-negative and asymptotically stable if and only if:
\begin{equation}
R_{d} =\frac{\beta _{1} \beta _{2} }{\alpha _{1} \alpha _{2} }  < 1
\end{equation}
and the equilibrium is otherwise unstable. The threshold $R_{d} $ is the \textit{basic dyadic relationship threshold}.  It says that the system is stable if product of the ratios of reactiveness and oblivious coefficients is less than one.
\end{thm}

\textit{Proof.}
Since the system is linear and two dimensional, we can study its Jacobian.  The Jacobian of the system (Eq. 1a-b) is given as:

\begin{equation}
J=\left[\begin{array}{cc} {-\alpha _{1} } & {\beta _{1} } \\ {\beta _{2} } & {-\alpha _{2} } \end{array}\right]
\end{equation}

Since the trace $\tau \left(J\right)=-\left(\alpha _{1} +\alpha _{2} \right)\le 0$ is non-positive, it suffices to show that the determinant $\Delta \left(J\right)$ is non-negative.  The determinant $\Delta \left(J\right)$ condition is given as:

\begin{equation}
\Delta \left(J\right)=\alpha _{1} \alpha _{2} -\beta _{1} \beta _{2}>0
\end{equation}

From this, we obtain the necessary and sufficient condition for stability and positivity of the equilibrium:

\begin{equation}
R_{d} =\frac{\beta _{1} \beta _{2} }{\alpha _{1} \alpha _{2} } < 1
\end{equation}

Equation $5$ is equivalent to equation $8$.  Note that if Eq. $7$ or equivalently equation $8$ is not satisfied, the equilibrium point is a saddle point, which is unstable and the positivity of equilibrium point does not hold. \qed

We can interpret theorem 1 as follows: for asymptotic stability, the squared geometric mean of the ratio of reactiveness to love and oblivion must be less than 1.  Whenever this statement does not hold, \textit{Strogatzian} model gives rise to unbounded feeling, which is obviously unrealistic.  For that reason, we restrict our study to the stable condition and state the following corollary for the linear model.

\begin{figure}
\begin{center}
\includegraphics[height=10cm]{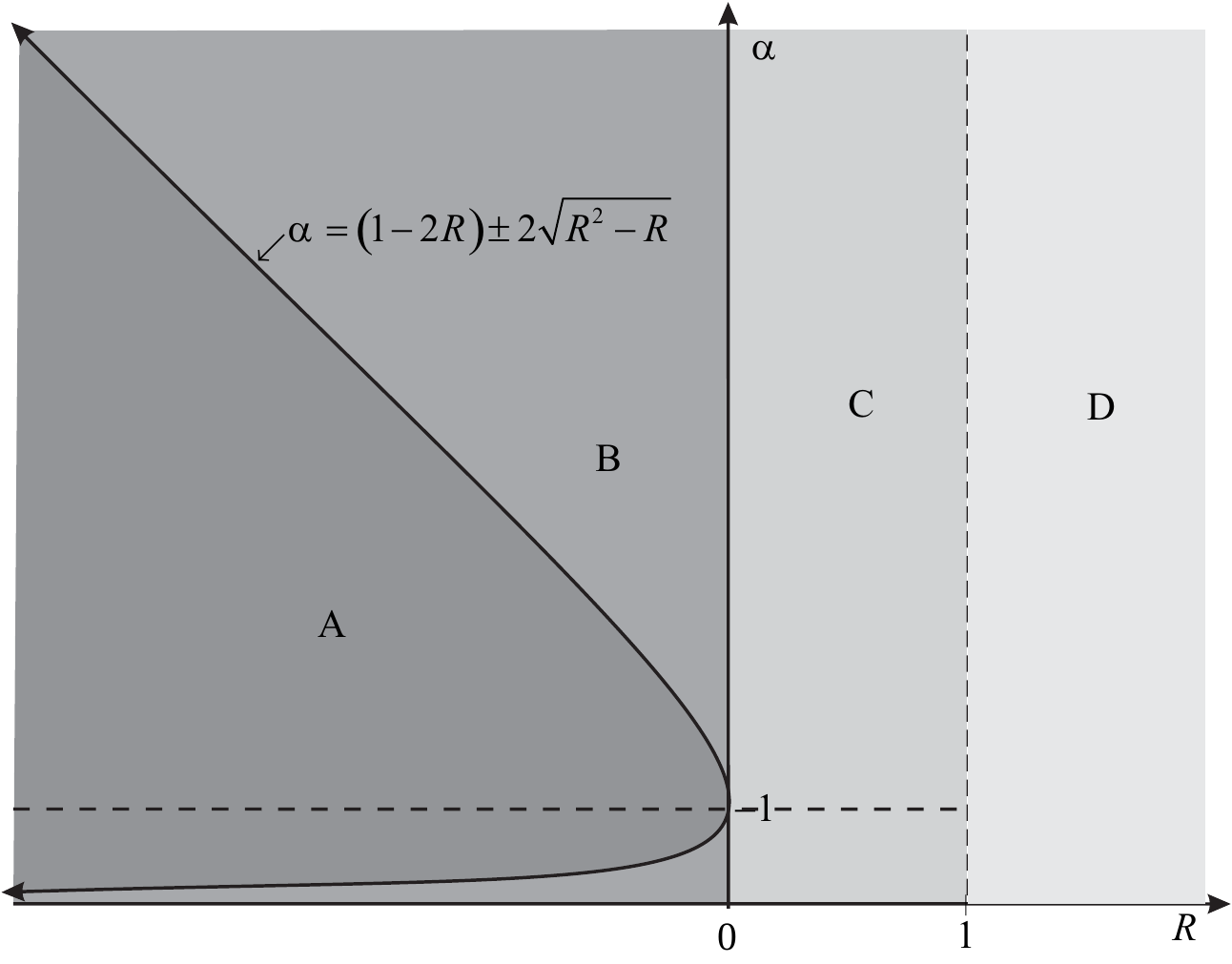}    
\caption{Stability Domain. Stable dynamics are observed in regions \textit{A}, \textit{B}, and \textit{C}, while unstable and unbounded trajectories reside in region \textit{D}}  
\label{fig1}                                 
\end{center}                                 
\end{figure}

\begin{corollary}
For the Eq. 1, and $\varepsilon =0$, we have the following:
\begin{itemize}
\item[(i)]  If $R_{d} <1$ and $\beta _{1} \beta _{2} >0$ or equivalently $0 < R_{d} <1$, then the equilibrium point of the system does not admit stable focus point or center.  The transients of $X_{i} \left(t\right)$ cannot have damped oscillations or other cyclic dynamics.
\item[(ii)]  If $R_{d} <1$ or $ R_{d} < 0$ and for some $\beta _{1} \beta _{2} <0$, then the equilibrium point admits stable focus.
\end{itemize}
\end{corollary}

 The results of Collorary 2.1.1 can be illustrated by the figure 2, where $\alpha$ is the ratio of $\alpha_{1}$ and $\alpha_{2}$ (e.g. $\alpha$ = $\frac{\alpha_{1}}{\alpha_{2}}$), and $R$ is the dyadic relationship threshold defined in equations $5$. In region \textit{D}, all equilibria are unstable and the dynamics give rise to unbounded (unrealistic) romantic feelings. Dynamics in the remaining regions (regions \textit{A}-\textit{C}) are stables. In region \textit{A} and \textit{B}, and \textit{C}, we obtain stable focus and node, respectively.  Later, we focus on the dynamics in region \textit{A} since their stochastic counterparts exhibit different dynamics for some parameter ranges. Figure 3 shows the phase portrait and time series of romantic dynamics residing in region \textit{A} in which the dynamics represent romantic behavior of secure and hermit individuals.

\begin{figure}
\begin{center}
\includegraphics[height=8.5cm]{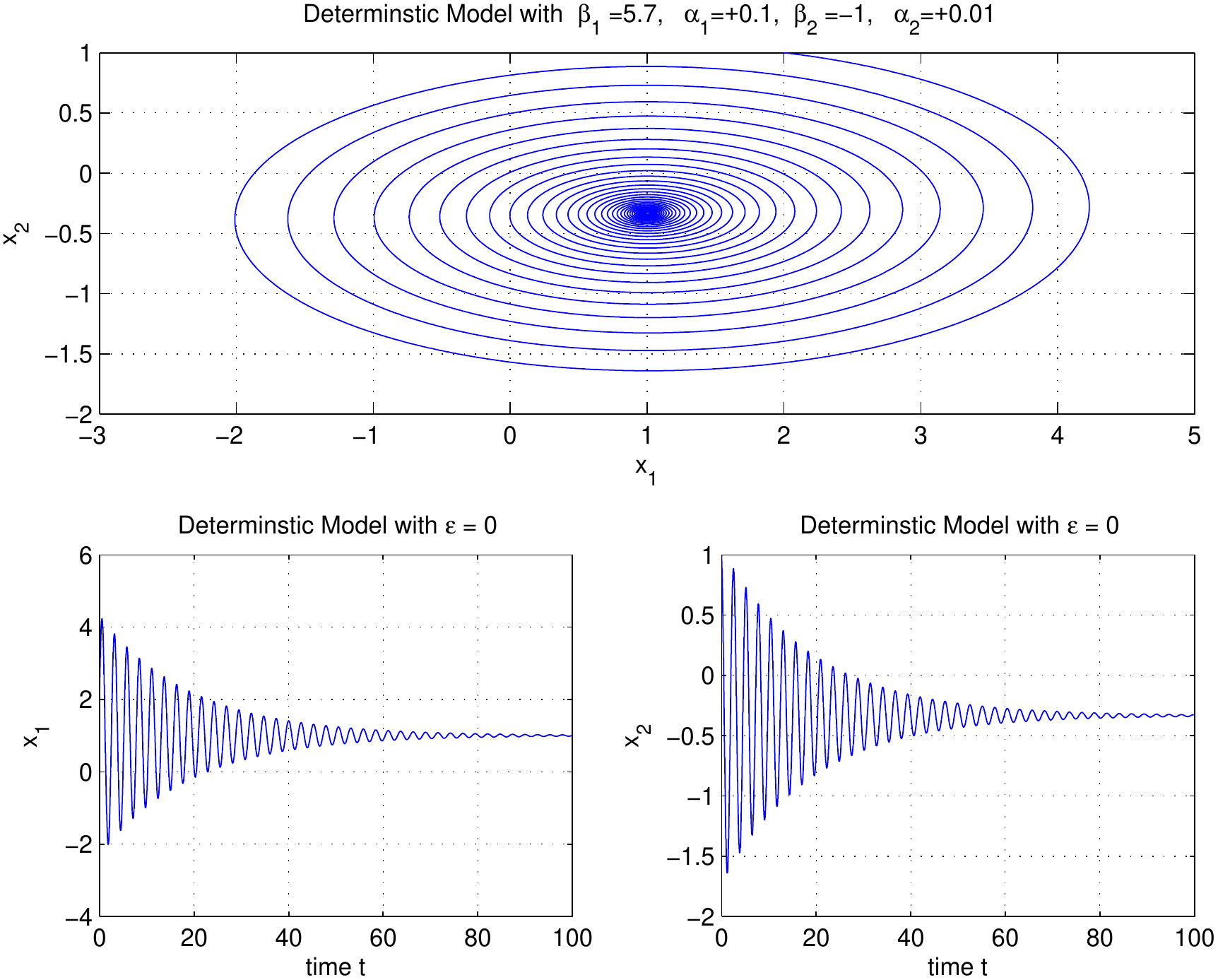}    
\caption{shows a deterministic evolution of love measure for a romantic relationship between secured or cautious and hermit lovers. Top plot shows the phase portrait of equation 1-2 with $\varepsilon = 0$, $\alpha_{1} = 0.1$, $\alpha_{2} = 0.01$, $\beta_{1} = 5.7$, and $\beta_{2} = -1$, linearized near a stable equilibrium. The bottom panel shows the time series solutions of romantic feelings, which exhibit damped oscillations}  
\label{fig1}                                 
\end{center}                                 
\end{figure}

For the nonlinear model where $\varepsilon$ $\neq$ $0$, the collorary $2.1.1$ does not hold entirely and theorem $1$ has minor correction term.  For more general models including the nonlinear case, the following statement holds.

\begin{thm}\label{theorem2}
The equilibrium $\left(\bar{X}_{1} ,\bar{X}_{2} \right)$ is asymptotically stable if and only if:

\begin{equation}
R_{d} =\frac{\beta _{1} \beta _{2} }{\alpha _{1} \alpha _{2} } d_{1} d_{2}  < 1
\end{equation}

where the correction term $d_{j} =\frac{dg\left(\bar{X}_{j} \right)}{dX_{j} } $ with $j=1,2$ and $g\left(u\right)$ is the linearized return function. The equilibrium is otherwise unstable.
\end{thm}

The proof of theorem $2$ is similar to that of theorem $1$. That is, we linearize around the steady state and change variables to obtain linear equations which then give rise to equation $9$ (see \cite{Liao2007} for a similar result).

 In the case of stochastic dynamics, our emphasis is placed on the dynamics that exhibit damped oscillations in the deterministic models (e.g. regions \textit{A} and \textit{B}), for the stochastic models fairly follow the deterministic behaviors of other equilibrium-type (e.g. stable nodes and limit cycle and/or centers) in region \textit{C}.  In the next section, we provide a stochastic dynamical system approach to illustrate some interesting dynamics observed in interpersonal and romantic relationships.

\subsection{Stochastic Dynamical Models}

 In the previous mathematical papers on the subject, all factors of relationship are independent of each other and they consider time-invariant personalities and the appeal of individuals, ignoring long-term aging, learning, adaptation processes, fast fluctuation of feelings, and external forces and influences such as familial approval and disapproval of loved ones. The role of oxytocin or vasopressin in the behavioral features, cultural and institutional conditions, and attachment dynamics are also ignored.  Accumulating all these forces as external factors that play major role on the quality of relationships, we consider stochastic variation of previous models (\cite{Cherif2009}) to investigate the stochastic nature of romantic dynamics by considering the rates as stochastic rates.  In the following section, we provide a method of using a deterministic formulation to derive a stochastic model.

\subsubsection{Derivation of Stochastic Love Dynamics}

In \cite{Cherif2009}, we outlined a similar method of deriving a stochastic equivalent of deterministic model.  The method used to derive the stochastic differential equations for dynamical process naturally lead to Ito stochastic differential equations, as oppose to other stochastic calculi (e.g. Stratonovich).  This paper only summarizes the approximation procedures, and as before interested readers should consult the work of Kurtz (\cite{Kurtz1971}) for more detailed treatment of the diffusion equation approximation, which corresponds to a continuous time Markov process.  We obtain the stochastic dynamical model for the processes by:

\begin{itemize}
\item[(i)] listing all the possible changes ${\rm \Delta X}=\left[\Delta X_{1} ,\Delta X_{2} \right]$ along with the probabilities for each change in a short time step $\Delta t$ (see table 1);
\item[(ii)] taking the expected changes $E\left[{\rm \Delta X}\right]$ and covariance matrix $E\left[{\rm \Delta X}\left({\rm \Delta X}\right)^{T} \right]$ are calculated for the Markov process.
\end{itemize}

Note that $E\left[{\rm \Delta X}\right]\left({\rm E}\left[{\rm \Delta X}\right]\right)^{T} =o\left({\rm \Delta t}^{{\rm 2}} \right)$ and can be ignored.  The rates in table 1 become the conditional transition rates of the stochastic process, that is, $P\left(X_{1,\left(t+\Delta t\right)} =x_{1} -1|X_{1} =x_{1} \right)=-\alpha _{1} X_{1} \Delta t+o\left(\Delta t\right)$ and so on.  To each of the increments, we add and subtract its conditional expectation, conditioned on the value of the process at the beginning of the time increment of length $\Delta t$.  This allows us to then decompose each increment into the sum of the expected value of the increment and sum of centered increment.  That is, $\Delta X_{1} =\left[-\alpha _{1} X_{1} +\beta _{2} X_{2} \left(1-\varepsilon X_{2} ^{2} \right)+A_{1} \right]\Delta t-\Delta Z_{1} +\Delta Z_{2} $ with the expected value of $E\left(\Delta X_{1} \right)=\left[-\alpha _{1} X_{1} +\beta _{2} X_{2} \left(1-\varepsilon X_{2} ^{2} \right)+A_{1} \right]\Delta t$, where the centered increment $\Delta X_{1} -E\left(\Delta X_{1} \right)$ is given as the difference of two increments, $\Delta Z_{2} -\Delta Z_{1} $.  The terms $\Delta Z_{i} $ are the difference of two centered Poisson increments.  These terms are then replaced by increment of Brownian motion $dW_{i} $ with corrected standard deviations or conditional variance.  The stochastic equations of the process can then be expressed in a form easily comparable to their deterministic equation counterpart.

\begin{table}
\begin{tabular}{|c|l|}\hline
\textbf{Transition} & \textbf{Rate} \\ \hline
$X_{1} \to X_{1} -1$ & $\alpha _{1} X_{1} $ \\ \hline
$X_{1} \to X_{1} +1$ & $\beta _{1} X_{2} \left(1-\varepsilon X_{2} ^{2} \right)+A_{1} $ \\ \hline
$X_{2} \to X_{2} -1$ & $\alpha _{2} X_{2} $ \\ \hline
$X_{2} \to X_{2} +1$ & $\beta _{2} X_{1} \left(1-\varepsilon X_{1} ^{2} \right)+A_{2} $ \\ \hline
\end{tabular}
\caption{Transition rate}
\end{table}

Alternatively, we can also arrive at the stochastic model by dividing the expected changes and the square root of the covariance matrix by $\Delta t$.  In the limit as $\Delta t\to 0$, the former becomes the drift term $\mu \left(t,X_{1} ,X_{2} \right)$, and the latter becomes the diffusion coefficient $D\left(t,X_{1} ,X_{2} \right)$, respectively.  Both procedures yield similar stochastic differential equations of the form:

\begin{equation}
dX=\mu \left(t,X_{1} ,X_{2} \right)dt+D\left(t,X_{1} ,X_{2} \right)dW
\end{equation}

where $W=\left[W_{1} ,...,W_{4} \right]^{T} $ is an independent Wiener process.  Notice that from the above formalism, the following statements are also true and can be verified:

\begin{equation}
E\left[\left|E\left(\frac{\Delta X}{\Delta t} \right)-\mu \left(t,X_{1} ,X_{2} \right)\right|^{2} \right]\to 0 as \Delta t\to 0
\end{equation}
and
\begin{equation}
E\left[\left|E\left(\frac{\Delta X\left(\Delta X\right)^{T} }{\Delta t} \right)-D\left(t,X_{1} ,X_{2} \right)D\left(t,X_{1} ,X_{2} \right)^{T} \right|^{2} \right]\to 0 as \Delta t\to 0
\end{equation}

The alternative procedure relies on using the discrete deterministic model and using similar argument as in the first method.  Using this approach, we arrive at a discrete stochastic model.  Conditions (Eqs. $11$-$12$) provide a justification for a weak approximation of moving from a discrete stochastic model to a continuous stochastic model.Eq.

This weak approximation is equivalent to the convergence of a family of discrete state-space Markov chains to a continuous stochastic process.  That is, for some class of smooth functions $G:\Re ^{2} \to \Re $ and let the solution to the stochastic differential equations be $X\left(T\right)$ at time T and the solution to the discrete stochastic equation be denoted by $X_{\Delta } \left(T\right)$, then $E\left[G\left(X\left(T\right)\right)\right]-E\left[G\left(X_{\Delta } \left(T\right)\right)\right]\to 0$ as $\Delta t\to 0$.  This provides a definition for weak convergence of discrete to continuous stochastic differential equations.  Kurtz (\cite{Kurtz1971}), and Kloden and Platen (\cite{Kloden1992})  have given detailed expositions on the methodology outlined in this section.

Using one of the methods sketched above, the stochastic differential equations describing the dynamics of romantic relationships are given as follows:
\begin{eqnarray}
dX_{1} =\left[-\alpha _{1} X_{1} +\beta _{1} X_{2} \left(1-\varepsilon X_{2} ^{2} \right)+A_{1} \right]dt-\sqrt{\alpha _{1} X_{1} } dW_{1} +\sqrt{\beta _{1} X_{2} \left(1-\varepsilon X_{2} ^{2} \right)+A_{1} } dW_{2}\\
dX_{2} =\left[-\alpha _{2} X_{2} +\beta _{2} X_{1} \left(1-\varepsilon X_{1} ^{2} \right)+A_{2} \right]dt+\sqrt{\beta _{2} X_{1} \left(1-\varepsilon X_{1} ^{2} \right)+A_{2} } dW_{3}-\sqrt{\alpha _{2} X_{2} } dW_{4}
\end{eqnarray}
for $\left(X_{1} ,X_{2} \right)\in R\times R$, where $W_{i} ,i=1,...,4$ are independent standard Wiener processes.  This approach allows us to extend deterministic models to stochastic models.  One can analyze the dynamics of the stochastic models with the help of the stability analysis of the deterministic equations.  In fact, the solution to the deterministic model corresponds to the mean of the stochastic model.  This framework provides a step towards understanding of the dynamics that are exhibited by the stochastic model.  It should be noted that the dynamics of the stochastic differential equations $13$-$14$ are closely related to the dynamics of the deterministic model (Eqs.~\ref{subeqn:first}-~\ref{subeqn:second}), but can exhibit important differences.  Additional dynamics can emerge in the stochastic model for some parameter values (Figs. $4$-$5$) for some $\beta _{1} \beta _{2} <0$ (e.g. for \textit{Strogatzian} model).  In (\cite{Cherif2009}), similar and more exotic behaviors were observed for nonlinear case and some of the results are included in herein, for both linear and nonlinear systems.

\noindent
\begin{center}
\begin{figure}
\begin{minipage}[b]{0.5\linewidth} 
\centering
\includegraphics[width=8.5cm]{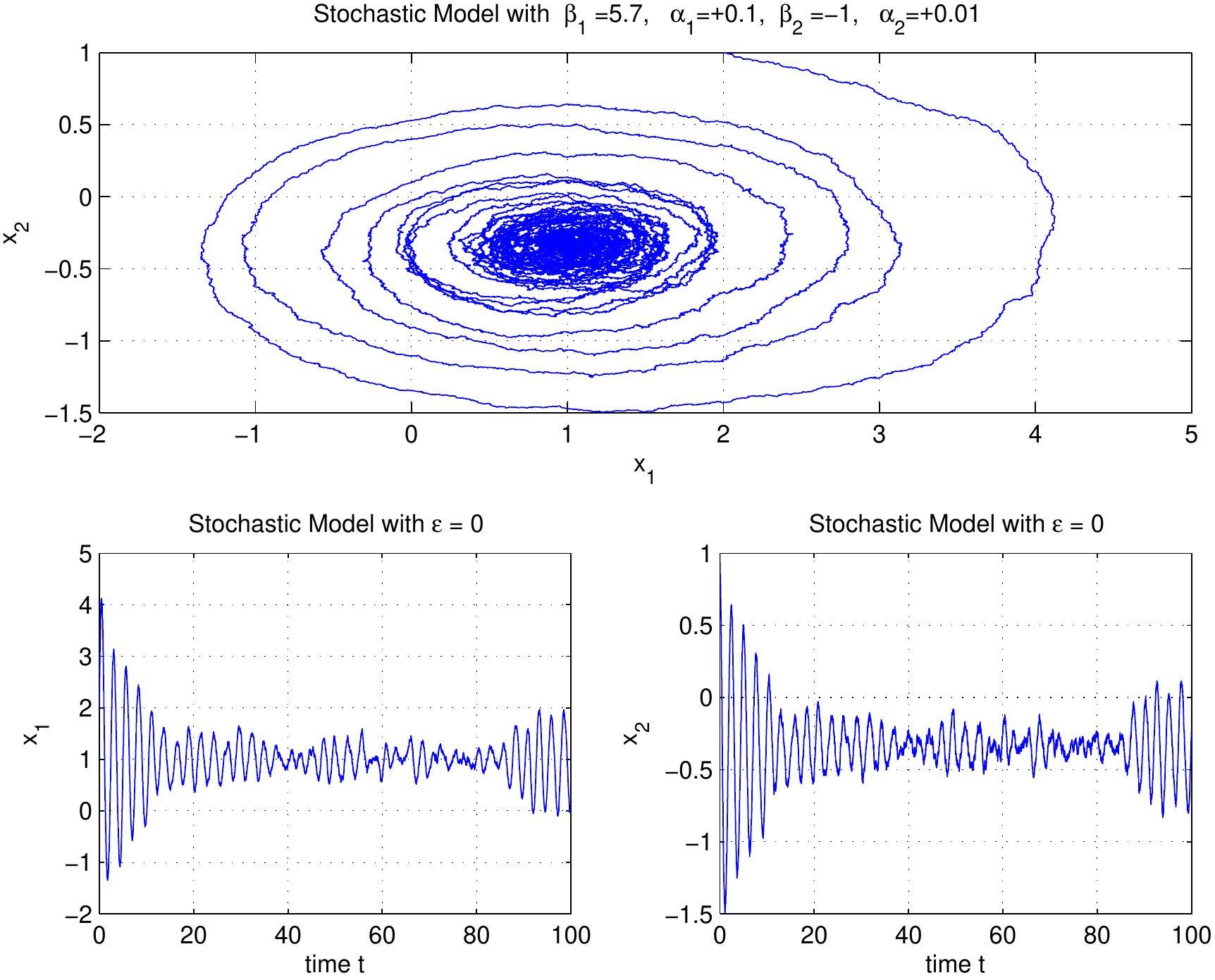}
\end{minipage}
\hspace{0.2cm} 
\begin{minipage}[b]{0.5\linewidth}
\begin{center}
\end{center}
\includegraphics[width=8.5cm]{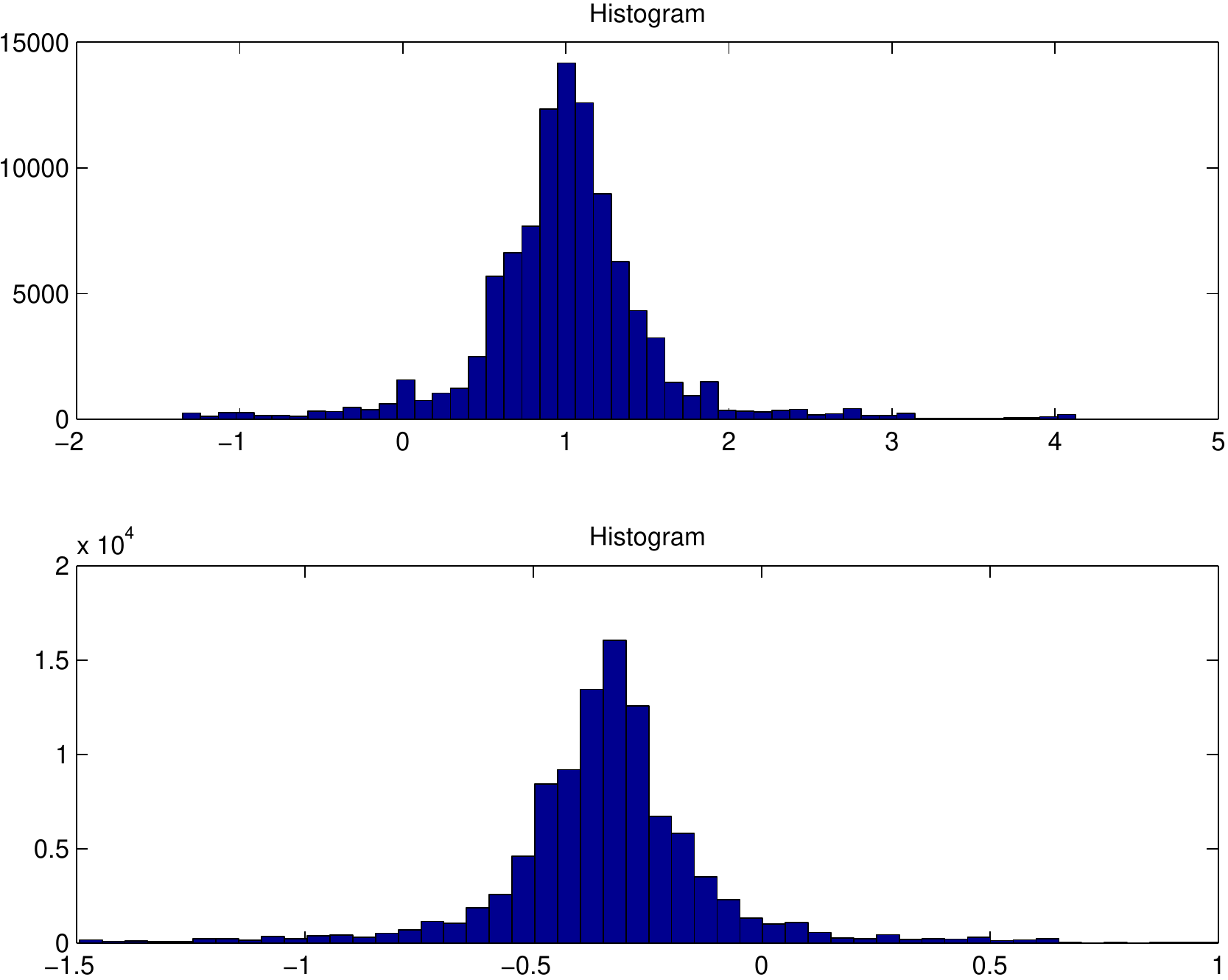}
\end{minipage}
\caption{shows the existence of sustained oscillation in the Stochastic model, while deterministic model does not exhibit such behaviors with the same parameter values $\varepsilon = 0$, $\alpha_{1} = 0.1$, $\alpha_{2} = 0.01$, $\beta_{1} = 5.7$, and $\beta_{2} = -1$. The deterministic system exhibits damped oscillations. It provides the contrast between the deterministic and stochastic models for the dynamics of romantic relationship. Left figure shows the distributions associated with the dynamics shown on the left.}
\end{figure}
\end{center}

For a stochastic \textit{Strogatzian} love affair (for the case where $\varepsilon = 0$), the system exhibits sustained oscillations whereas a deterministic model shows damped oscillation (see figure $4$). In the case of nonlinear model ($\varepsilon >0$) with appropriate conditions being satisfied, we observe various dynamics including those observed in the linear case (e.g. sustained oscillation via stochastic resonance).

\noindent
\begin{center}
\begin{figure}
\begin{minipage}[b]{0.5\linewidth} 
\centering
\includegraphics[width=8.5cm]{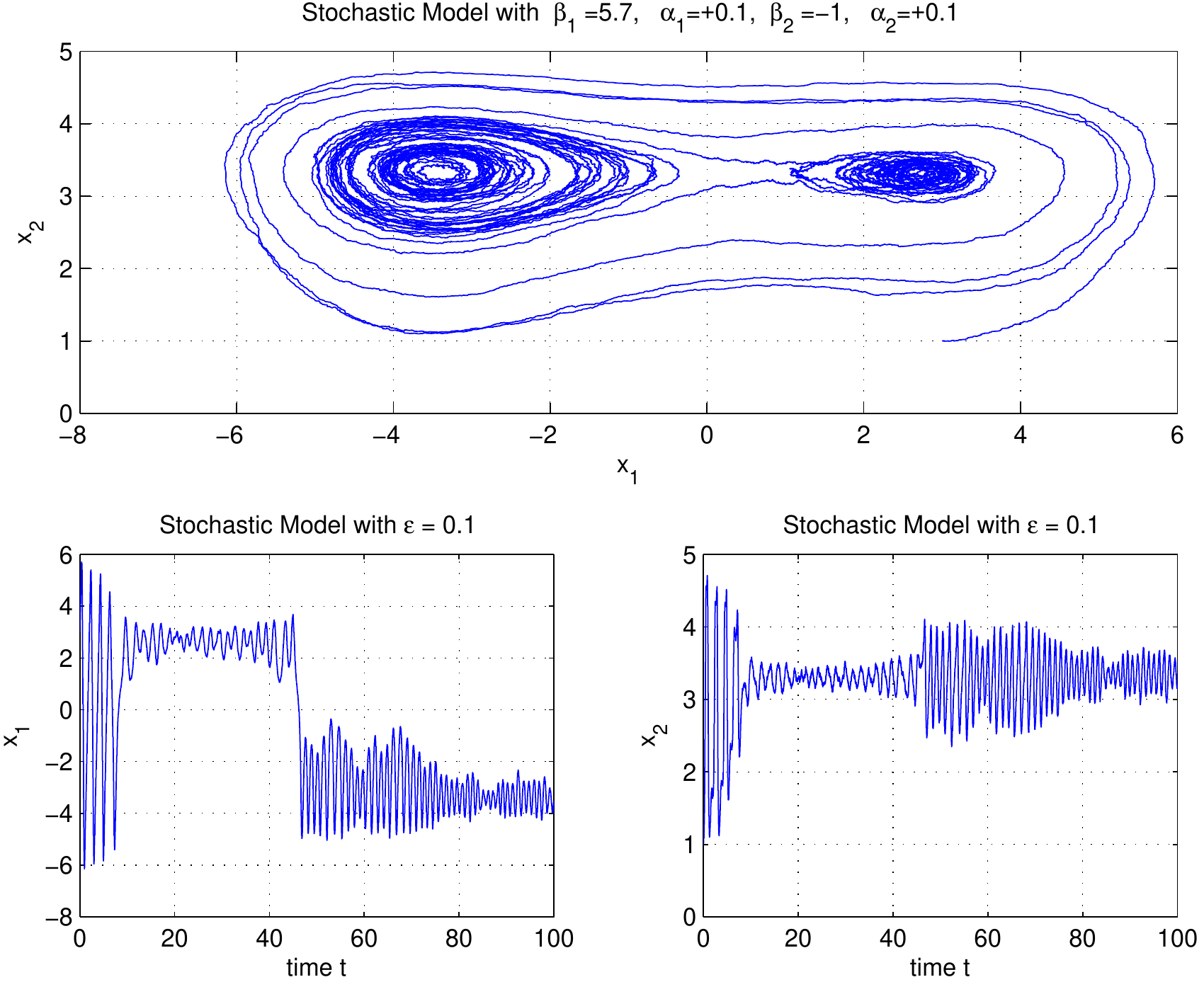}
\end{minipage}
\hspace{0.2cm} 
\begin{minipage}[b]{0.5\linewidth}
\begin{center}
\end{center}
\includegraphics[width=8.5cm]{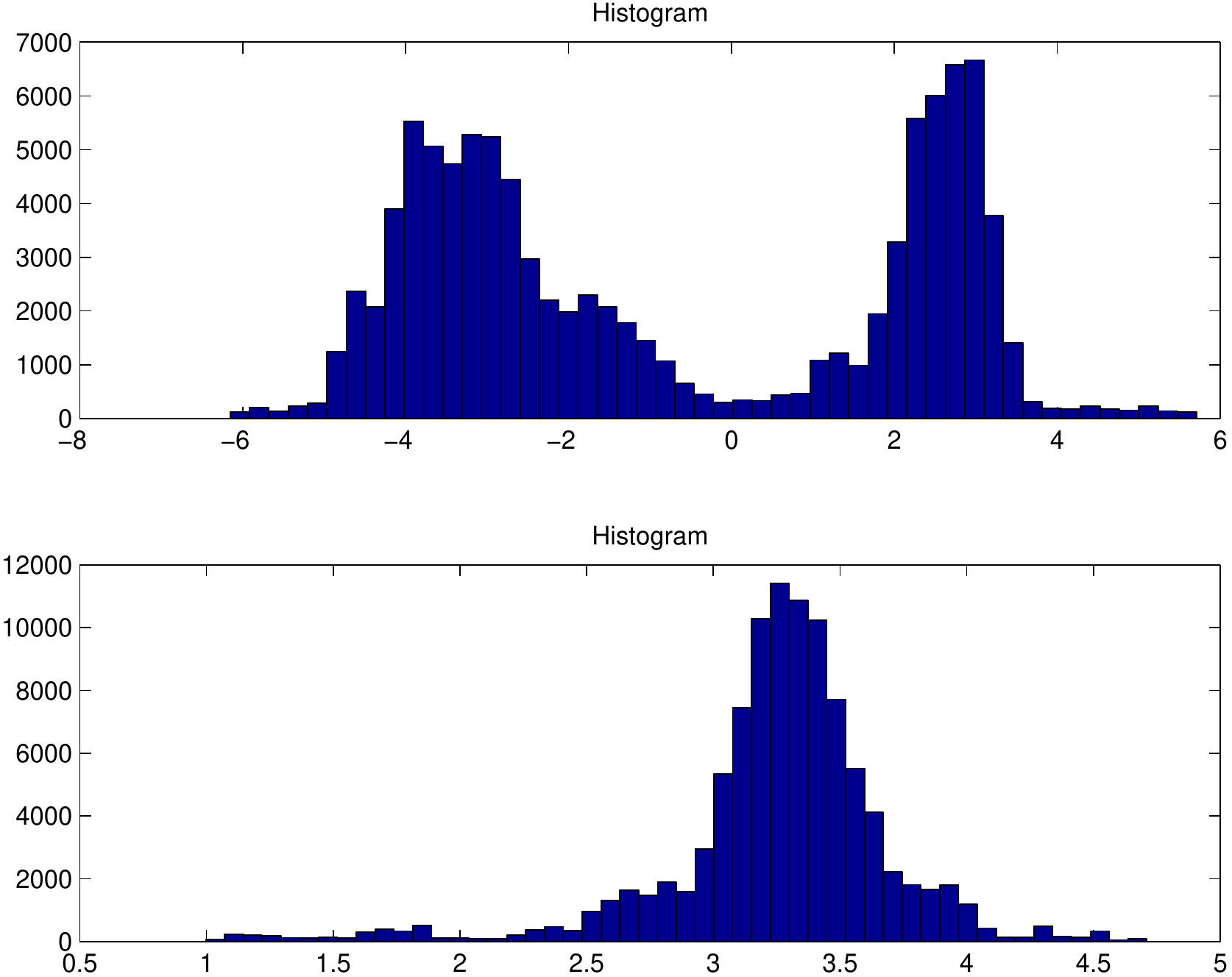}
\end{minipage}
\caption{shows diffusion between two locally stable equilibria in stochastic dynamics of love affair. It corresponds to the nonlinear stochastic dynamics with similar $\beta_{i}$ parameter values as in figure 3, where we use the following parameters: $\varepsilon \neq 0$, $\alpha_{1} = \alpha_{2} = 0.1$, $\beta_{1} = 5.7$, and $\beta_{2} = -1$. The distribution for $X_{1}$ is bi-modal while that of $X_{2}$ is unimodal.}
\end{figure}
\end{center}

In figure $5$, oscillations can persist or are sustained for some parameter values, whereas the deterministic equations exhibit damped oscillations (figures $4$ and $5$). Note that in figure $5$, we observed both sustained oscillations at the equilibria and ``jumping'' or ``switching'' phenomena.  In the case of ``switching,'' there are trajectories which diffuse from equilibrium to equilibrium (see figure $5$).  A more detailed investigation of such dynamics (e.g. sustained and ``jumping'' oscillations or transition between locally stable equilibria), we hope, will be studied in the next paper, where a modification of multiple time-scale approach with Ito-Doeblin Formula (also known as Ito formula) can be employed.  The system exhibiting sustained oscillations with transitions between local stable equilibria is only observed in fragile interpersonal and romantic relationships (figure $5$), while robust interpersonal and romantic relationships shows only sustained oscillations (figure $4$).  These behavioral dynamics are dependent on the variance associated with the diffusion terms, which can easily be verified with multiple time-scale method outlined by Kuske \textit{et al} (\cite{Kuske2007}).  We also observed (not show on the figures) that for some parameter values, the transitions (jump) between equilibria or sustained oscillations are transient and are not sustained for these values.  In other cases, for different conditions, the trajectories visit most of the equilibria of the system in a sustained way. For example, when there are more than two locally stable equilibria satisfying corollary 2.1(\textit{ii}) condition, the trajectories can visit most (if not all) of the steady states in fragile relationships, hence exhibiting multi-modal distributions.  Therefore, this paper illustrates the need for more mathematical analysis of interpersonal and romantic relationships from the perspective of nonlinear stochastic differential equations.

\subsubsection{Extension: Differential Romantic Style}
Using similar methods as outlined above, a more complex stochastic dynamic of love affairs can be proposed. The ansatz is as follows:
\begin{eqnarray}\label{ee}
dX_{1} =\left[-\alpha _{1} X_{1} +\beta _{1} X_{2} \left(1-\varepsilon X_{2} ^{2} \right)+A_{1} \right]dt-\sqrt{\alpha _{1} X_{1} } dW_{1} +\sqrt{\beta _{1} X_{2} \left(1-\varepsilon X_{2} ^{2} \right)+A_{1} } dW_{2}\\
dX_{2} =\left[-\alpha _{2} X_{2} +\beta _{2} X_{1} \left(1-\varepsilon X_{1} ^{2} \right)+A_{2} \right]dt+\sqrt{\beta _{2} X_{1} \left(1-\varepsilon X_{1} ^{2} \right)+A_{2} } dW_{3}-\sqrt{\alpha _{2} X_{2} } dW_{4}\\
d\alpha_{1} = \alpha_{10}\left[\alpha_{11} - \alpha_{1} \right]dt + \sqrt{\alpha_{12}}dW_{5}\\
d\alpha_{2} =\alpha_{20}\left[\alpha_{21} - \alpha_{2} \right]dt + \sqrt{\alpha_{22}}dW_{6}\\
d\beta_{1} =\beta_{10}\left[\beta_{11} - \beta_{1} \right]dt + \sqrt{\beta_{12}}dW_{7}\\
d\beta_{2} =\beta_{20}\left[\beta_{21} - \beta_{2} \right]dt + \sqrt{\beta_{22}}dW_{8}\\
\end{eqnarray}

for $\left(X_{1} ,X_{2}, \alpha_{1}, \alpha_{2}, \beta_{1}, \beta_{2}  \right)\in R^{2}\times R^{4}$, where $W_{i} ,i=1,...,8$ are independent standard Wiener processes. Equations ($15$-$21$) represent a stochastic model of love affair with variability in romantic style which is impacted by environmental factors such as dislike of family members and/or friends against one's partner, ecological and institutional conditions, bio-sociological factors, etc... Variations due to these factors can also affect the measure of love in an unpredictable matter. One way to capture these variations is to have different variables for each one of the factors of interest. However, this will complicate the analysis, and render the analysis of the system intractable. The equations above simplify and assume that, for example,  $\beta_{1}(t,\omega_{1}, ..., \omega_{n})$, where  $\omega_{i}$, $i=1,..,n$ , are different romantic factors that affect the romantic style of individuals. Note that each of these parameter dynamics $\alpha_{1}$, $\alpha_{2}$, $\beta_{1}$, and $\beta_{2}$ can be integrated and solved exactly. We can then take their limit cases (steady state conditions) and substitute them in equations $15$-$16$ to reduce our system to equations $13$-$14$. Assuming that romantic styles vary stochastically, one can also investigate the possible effect of the inclusion of these dynamics.  One such effect is the spread in distribution of evolution of feelings. However, this analysis is left to the next paper and was not investigate herein.

\section{Conclusion}
In this paper, we have considered both deterministic and stochastic models with nonlinear return functions. The stochastic model has a structure related to a deterministic model which allows us to study most of its dynamics through the lens of deterministic analysis. We have focused on particular subsets of interesting dynamics that are not observed in deterministic models. While a deterministic model exhibited damped oscillations with certain parameter values, the stochastic models showed sustained oscillations with the same parameter values. The results show that deterministic linear and nonlinear models tend to approach locally stable emotional behaviors. However, in the presence of stochasticity in the models, more complex and exotic patterns of emotional behaviors are observed. The stochastic differential equation extension provides insight into the dynamics of romantic relationships that are not captured by deterministic models, which assumes that love is scalar and individuals respond predictably to their feelings and that of others without external influences, such as ecological factors. Stochastic models capture the fluctuations due to the dynamics of love and that of external influences. This paper provides a new direction to the study of interpersonal relationship.  The future direction toward more realistic mathematical and theoretical modeling of the dynamics of romantic relationship is possible through the lens of agent-based modeling, where community interaction is included, or integrative models (interpersonal, bio-sociological and ecological models are integrated) can be investigated. Calculation of the distribution of time indifference or apathy of individuals is a new possibility arising out of the stochastic dynamics of romantic relationships.  The most fruitful direction from mathematical purview is developing methods to analysis systems that exhibit "stability boundary crossing" or "jump between locally stable equilibria" dynamics.  The analysis of these dynamics, to the best of our knowledge, has not been investigated elsewhere and is worth studying in another paper.

\begin{ack}
The authors acknowledge the financial supports of Mathematical, Computational and Modeling Sciences Center (MCMSC), which is supported in part by the National Science Foundation (NSF) under the grant number DMS-0502349, the National Security Agency (DOD-H982300710096), and the Sloan Foundation.  One of the authors (A. Cherif) is also supported in part by NSF under the auspice of Cohort VI WAESO LSAMP Bridge to the Doctorate Fellowship, the Sloan Fellowship and the ASU Graduate College Diversity Enrichment Fellowship. The authors thank Dr. Priscilla Greenwood, Dr. Marco Janssen, Dr. Carlos Castillo-Chavez and Dr. Stephen Tennenbaum for their suggestions and comments.
\end{ack}

\bibliography{autosam}           

\end{document}